\documentclass[journal=acsphotonics,manuscript=article]{achemso}

\usepackage{chemformula} 
\usepackage[T1]{fontenc} 

\usepackage[colorlinks,citecolor=blue,linkcolor=red]{hyperref}
\usepackage{graphicx}
\usepackage{subfigure}
\usepackage{epsfig}
\usepackage{dcolumn}
\usepackage{bm}
\usepackage{times}
\usepackage{amsmath}
\usepackage{float}
\usepackage{multirow}
\usepackage[normalem]{ulem}
\usepackage{color}

\author{Jiebin Peng}
\affiliation{Center for Phononics and Thermal Energy Science, China-EU Joint Center for Nanophononics, Shanghai Key Laboratory of Special Artificial Microstructure Materials and Technology, School of Physics Science and Engineering, Tongji University, 200092 Shanghai, China}
\author{Gaomin Tang}
\email{gaomin.tang@unibas.ch}
\affiliation{Department of Physics, University of Basel, Klingelbergstrasse 82, CH-4056 Basel, Switzerland}
\author{Luqin Wang}
\affiliation{Center for Phononics and Thermal Energy Science, China-EU Joint Center for Nanophononics, Shanghai Key Laboratory of Special Artificial Microstructure Materials and Technology, School of Physics Science and Engineering, Tongji University, 200092 Shanghai, China}
\author{Rair Mac\^{e}do}
\affiliation{James Watt School of Engineering, Electronics \& Nanoscale Engineering Division, University of Glasgow, Glasgow G128QQ, United Kingdom}
\author{Hong Chen}
\affiliation{Center for Phononics and Thermal Energy Science, China-EU Joint Center for Nanophononics, Shanghai Key Laboratory of Special Artificial Microstructure Materials and Technology, School of Physics Science and Engineering, Tongji University, 200092 Shanghai,
China}
\author{Jie Ren}
\email{xonics@tongji.edu.cn}
\affiliation{Center for Phononics and Thermal Energy Science, China-EU Joint Center for Nanophononics, Shanghai Key Laboratory of Special Artificial Microstructure Materials and Technology, School of Physics Science and Engineering, Tongji University, 200092 Shanghai,
China}

\title[An \textsf{achemso} demo]
  {Twist-induced Near-field Thermal Switch Using Nonreciprocal Surface Magnon-Polaritons}

\abbreviations{SMP,FMI}

\begin{document}

\begin{tocentry}
 Through the ‘twist angle’ between two ferromagnetic insulators, we introduce a large near-field thermal switch and the nonmonotonic twist manipulation for heat flux.
\end{tocentry}

\begin{abstract}
We explore that two ferromagnetic insulator slabs host a strong twist-induced near-field
radiative heat transfer in the presence of twisted magnetic fields. Using the formalism of
fluctuational electrodynamics, we find the existence of large twist-induced thermal switch
ratio in large damping condition and nonmonotonic twist manipulation for heat transfer in
small damping condition, associated with the different twist-induced effects of
nonreciprocal elliptic surface magnon-polaritons, hyperbolic surface magnon-polaritons, and
twist-non-resonant surface magnon-polaritons. Moreover, the near-field radiative heat transfer can
be significantly enhanced by the twist-non-resonant surface magnon-polaritons in the ultra-small damping
condition. Such twist-induced effect is applicable for other kinds of anisotropic slabs
with time-reversal symmetry breaking. Our findings provide a way to twisted and magnetic
control in nanoscale thermal management and improve it with twistronics concepts.
\end{abstract}

\bigskip

A key component for manipulating radiative heat flow at the nanoscale is near-field
radiative heat transfer, which can exceed Planck's blackbody radiation limit~\cite{Planck}
by orders of magnitude due to the presence of evanescent modes~\cite{thermal-APR,
PvH, review07, review15, review18, review15-2,Kim2015,Cui2017,JS1,gm1,gm2}. Two types of
surface modes have been commonly studied in near-field heat transfer; one is surface
plasmon-polaritons~\cite{SPP1,SPP2,SPP3,SPP-graphene1,SPP-graphene2,SPP-graphene3,SPP-graphene4,SPP-BP,SPP-Si1,SPP-Si2, SPP-Si19} and the other is surface phonon-polaritons~\cite{SPhP1,SPhP2,SPhP3,SPhP4,SPhP5,SPhP6}. In addition, surface magnon-polaritons (SMPs), hybrid collective excitations due to the coupling between magnons and
electromagnetic fields~\cite{MATSUURA1983157,disp-FM-85,PhysRevBMac}, also has functional
associations to thermal management in nanotechnologies.
For instance, in magnetic recording devices, a magnetic read/write head touches above the disk surface with nanometers
separation.
At such a short distance, SMPs should play a significant role in the near-field thermal
manipulation of magnetic recording devices.
Moreover, due to the high gyrotropic optical effect~\cite{Polder49}, SMPs in uniaxial
ferromagnetic insulator (FMI) are nonreciprocal.
Such nonreciprocal behavior can break Kirchhoff's law~\cite{Miller4336} and paves the way
for the exploitation of radiative thermal transfer at nanoscale.

Recently, twistronics becomes an emerging research topic since the electronic state can be
manipulated through the ``twist angle" between two layers, leading to flat-band
superconductivity~\cite{Unconventional2018,Strange2020}, moir\'{e}
excitons~\cite{2017Moir}, stacking-dependent interlayer magnetism~\cite{0Direct} and other
exotic electronic properties. Similar twist-induced concepts have been demonstrated in
photonics, such as moir\'{e} photonics crystal~\cite{2018Photonic}, moir\'{e} hyperbolic
metasurfaces~\cite{2020Moir} and photonic magic
angles~\cite{2015Hyperbolic,Topological2020}. Motivated by these exotic discoveries,
several works have shown the development of tunable radiative heat flow between
two-dimensional materials and biaxial
crystals~\cite{2018Influence,He_20,2020Near,2020Near_field,Polariton_2020} through twist.
With the analogous principle, we explore the effects of radiative thermal twistronics
between the uniaxial FMIs with external magnetic fields, where the twist and nonreciprocal
phenomena can both arise in the domain of thermal management.

In this Letter, we consider to manipulate near-field radiative heat transfer through the twist between two uniaxial FMIs. Nonreciprocal SMPs emerge at the interface between
vacuum and gyrotropic FMIs with asymmetric permeability tensor. Based on the
nonreciprocity, we demonstrate a large twist-induced near-field thermal switch effect with
a moderate external magnetic field. Under ultra-small damping condition, we show an unusual
twist-induced near-field thermal transfer enhancement due to the presence of
twist-non-resonant SMPs.


\begin{figure}
\centering
  \includegraphics[width=\columnwidth]{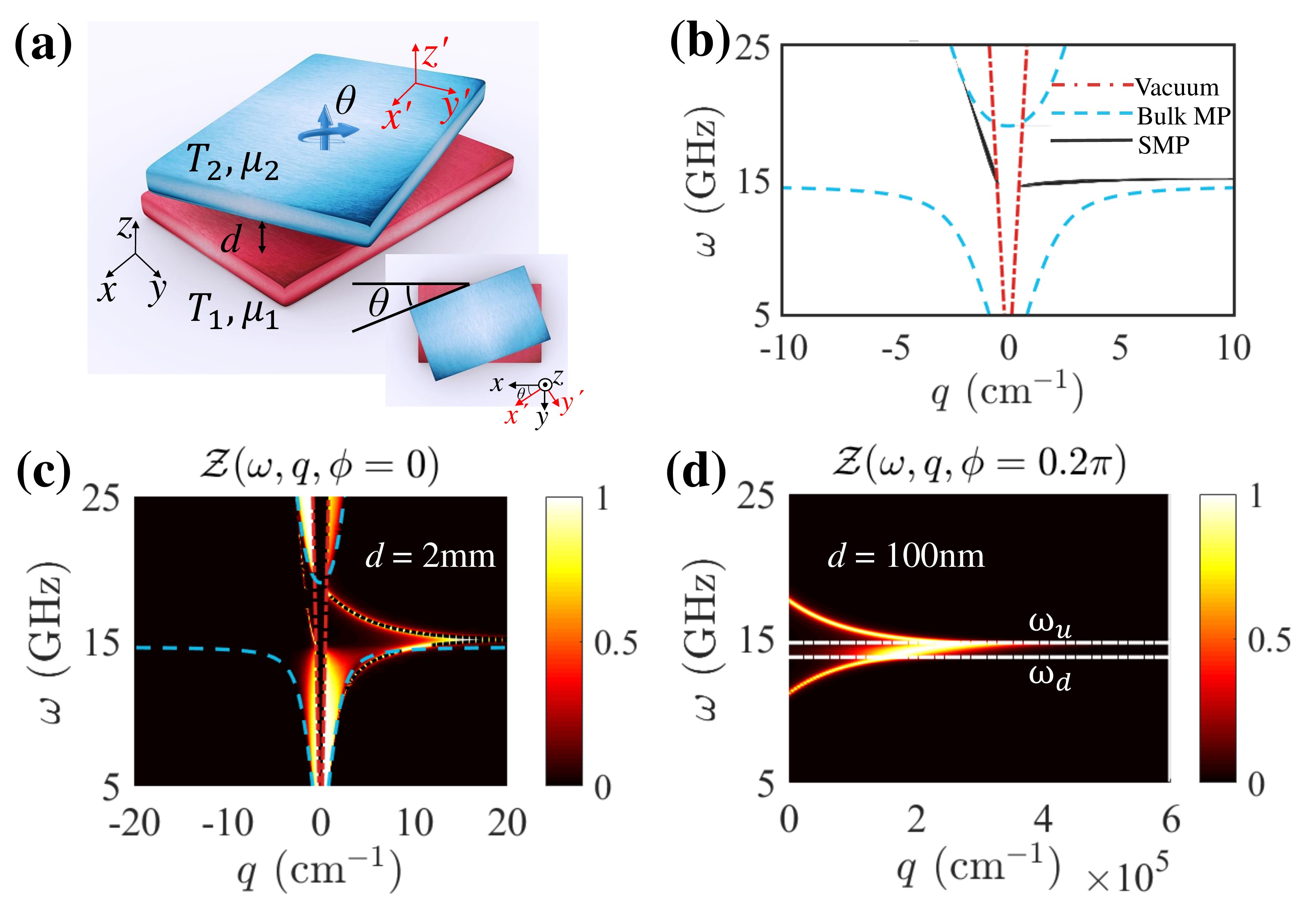} \\
  \caption{(a) A schematic setup for radiative heat transfer between two FMIs with vacuum
    separation $d$. The bottom and top slabs have the temperature $T_1$ and $T_2$,
    respectively. The $y$ ($y'$) axis is along the direction of the satuation
    magnetisation in the bottom (top) FMI. The magnetic fields in each slab is applied
    along the direction of the corresponding satuation magnetization.
    The twist angle $\theta$ is defined by the anticlockwise rotation of $x'y'z'$
    coordinate system with respect to $xyz$ coordinate system.
    (b) Dispersion relation of nonreciprocal SMP with a single vacuum-FMI interface.
    (c) Energy transmission coefficient ${\cal Z}(\omega,q,\phi=0)$ with gap distance
    $d=2\,$mm. The cyan dashed line and the red dash-dotted line are the same as in (b).
    The black dotted line shows the nonreciprocal symmetric and asymmetric modes of SMPs.
    (d) Energy transmission coefficient ${\cal Z}(\omega,q,\phi=0.2\pi)$ with gap distance
    $d=100\,$nm. The damping
    constant $\alpha$ is $0.01$ in (c) and (d). }
  \label{fig1}
\end{figure}
{\it Radiative heat transfer.--}
We consider near-field radiative heat transfer between two FMIs with temperatures
$T_{1(2)}= T\pm \Delta T/2$ and twist angle $\theta$ [See Fig.~\ref{fig1}(a)]. A Cartesian
coordinate system $xyz$ ($x'y'z'$) is defined at the bottom (top) slab and the $y$ ($y'$)
axis is along the direction of the applied magnetic field and saturation magnetisation.
The twist angle $\theta$ is defined as the angle between the $y'$ and $y$ axis.
We define the heat transfer coefficient $\kappa$ as $\kappa =\lim_{\Delta T \rightarrow 0}
J / \Delta T$ with $J$ the heat flux.
From fluctuational electrodynamics~\cite{PvH, review07}, the heat transfer coefficient can
be expressed as
\begin{align} \label{HTC_DEF}
  \kappa(T,\theta) =\int_0^{\infty} \frac{d\omega}{2\pi} \hbar\omega \frac{\partial
  N}{\partial T} \int_{0}^{\infty} \frac{dq}{2\pi} q \int_{0}^{2\pi} \frac{d\phi}{2\pi}
  Z(\omega,q,\phi),
\end{align}
where $q$ is the in-plane wave vector and $\phi$ the in-plane azimuthal angle.
In the above expression, $\partial N/\partial T$ is the derivative of the Bose
distribution function with respect to the temperature. We consider the relative heat
transfer coefficient scaled by the black-body limit $\kappa_{b} = 4\sigma_b T^3$ with
$\sigma_b = \pi^2 k_B^4/(60 \hbar^3 c^2)$. The energy transmission coefficient ${\cal
Z}(\omega,q,\phi)$ with twist angle $\theta$ reads
\begin{align}
  {\cal Z} = \left\{
  \begin{aligned}
  &{{\rm Tr}[({\bf I}-{\bf R}^\dagger_2{\bf R}_2) {\bf D} ({\bf I}-{\bf R}_1{\bf R}^\dagger_1) {\bf
  D}^\dagger] }, & q < \omega/c , \\
  &{{\rm Tr}[({\bf R}^\dagger_2 -{\bf R}_2) {\bf D} ({\bf R}_1 - {\bf R}^\dagger_1) {\bf D}^\dagger] }
  e^{-2|\beta_0|d}, & q > \omega/c,
\end{aligned}
\right.
\end{align}
where $\beta_0= \sqrt{(\omega/c)^2-q^2}$ is the out-of-plane wave vector in vacuum and
${\bf I}$ the identity matrix. The Fabry-Perot-like denominator matrix is written as ${\bf
D}=({\bf I}-{\bf R}_1 {\bf R}_2 e^{2i\beta_0 d})^{-1}$. In our setup, the reflection
coefficient matrix ${\bf R}_a$ with $a=1,2$ is written as
\begin{equation}
  {\bf R}_a =
  \begin{bmatrix}
  r^a_{ss} & r^a_{sp} \\
  r^a_{ps} & r^a_{pp}
  \end{bmatrix}
\end{equation}
where superscripts $s$ and $p$ denote the polarization states. The reflection coefficients
can be calculated by the transfer matrix methods~\cite{Pochi1980Optics} and the details
are given in the Supplemental Material~\cite{SM}. For later convenience, we also define
the integrated energy transmission coefficient, i.e. ${\cal{Z}}(\omega,\phi)$, which is the energy transmission coefficient after an integration over the wave
vector $q$.

By applying a magnetic field along the $y$-direction in the bottom FMI, the permeability tensor has the form~\cite{Polder49, disp-FM-85}
\begin{equation}
  \mu =
  \begin{bmatrix}
  \mu_{xx} & \mu_{xy} & \mu_{xz} \\  \mu_{yz} & \mu_{yy} & \mu_{yz} \\ \mu_{zx} & \mu_{zy} & \mu_{zz}
  \end{bmatrix} =
  \begin{bmatrix}
  \mu_r & 0 & -i\mu_i \\  0 & 1 & 0 \\ i\mu_i & 0 & \mu_r
  \end{bmatrix} ,
\end{equation}
The diagonal and off-diagonal terms are, respectively, expressed as
$\mu_r=1+\frac{\omega_m(\omega_0+i\alpha\omega)}{(\omega_0+i\alpha\omega)^2-\omega^2}$ and
$\mu_i =\frac{\omega_m \omega}{(\omega_0+i\alpha\omega)^2-\omega^2}$ with frequency
$\omega$ and magnetic precession damping constant $\alpha$.
The magnetic resonance frequencies $\omega_0 = \mu_0\gamma h$ and $\omega_m = \mu_0\gamma m_s$ are
due to the external magnetic field $h$ and the saturation magnetization $m_s$ with the
gyromagnetic ratio $\gamma$.
The relative permittivity of the FMI is assumed to be a constant.
For the top FMI, the relative permeability tensor is expressed as $\mu' = {\cal R}(\theta) \mu {\cal R}^T(\theta)$ with the rotation matrix ${\cal R}(\theta)$ along $z$ axis.
During the numerical calculation, we adopt the parameters of yttrium iron garnet
(YIG) with the relative permittivity $\epsilon=14.5$~\cite{YIG16},
gyromagnetic ratio $\gamma/2\pi =28\,$GHz/T~\cite{Rair-Macdo2020} and saturation magnetization $\mu_0 m_s=0.28\,$T~\cite{Boventer2018}. The applied magnetic field $\mu_0 h$ is taken as $0.4\,$T. Such set of parameters results in SMPs at microwave frequency range so that we consider the radiative heat transfer at the cryogenic environment (around $4\,$K).

{\it Nonreciprocal surface magnon-polaritons.--}
At a single vacuum-FMI interface, there exists SMPs of which the dispersion is
nonreciprocal. The implicit dispersion relation for SMPs is~\cite{SM}
\begin{equation} \label{SMP_s}
  \beta_0 + (\mu_r \beta_1 - i\mu_i q)/(\mu_r^2-\mu_i^2)=0 .
\end{equation}
where $\beta_1=\sqrt{\epsilon \mu_{eff}(\omega/c)^2 -q^2}$ is the
out-of-plane wave vector inside the FMI and $\mu_{eff}=(\mu_r^2-\mu_i^2)/\mu_r$. Figure.~\ref{fig1}(b) indicates the
nonreciprocal dispersion of SMPs (gray line) outside the light cone (red
dash-dotted line), together with the symmetric dispersion of bulk magnon-polations, that is, $q=\sqrt{\epsilon \mu_{eff}} \omega/c$ (cyan dashed line). We highlight that SMPs exist at the band gap region of FMI and the
high-$q$ SMPs only exist at positive wave vector region, which is useful in manipulating
near-field heat transfer.

For the case of two FMIs with millimeter separation, SMPs from two interfaces can be
coupled. Figure.~\ref{fig1}(c) shows the energy transmission coefficient between two FMIs
at zero azimuthal and twist angle, that is, ${\cal Z}(\omega,q,\phi=0;\theta=0)$. We can
observe that there exists an asymmetric transmission coefficient both for bulk MPs (the
region outside the light cone and inside the dispersion relation of MPs) and SMPs (the
near-unity line inside the band gap), with respect to the in-plane wave vector.
The two near-unity lines for SMPs are consistent with the implicit dispersion relation of
SMPs as follows
\begin{align}
  & \beta_0 + \tanh(|\beta_0|d/2) (\beta_1 \mu_r - iq \mu_i)/(\mu_r^2 -\mu_i^2) = 0,
  \label{DR1} \\
  & \beta_0 + \coth(|\beta_0|d/2) (\beta_1 \mu_r - iq \mu_i)/(\mu_r^2 -\mu_i^2) = 0.
  \label{DR2}
\end{align}
In the absence of the contributions from $\mu_i$, Eqs.~\eqref{DR1} and \eqref{DR2} can be
reduced to dispersion relations similar to those of surface phonon-polaritons.

In addition, the optical properties of FMI are anisotropic in the $x$-$z$ plane when there
is nonzero azimuthal angle.
To qualitatively analyze the anisotropic effects, we show the energy transmission
coefficient with a nonzero azimuthal angle in Fig.~\ref{fig1}(d), where the near-unity
lines between frequency $\omega_{u}$ and $\omega_{d}$ expand as a near-unity spot.
Here, $\omega_{u}$ and $\omega_{d}$ are the $\mu$-near-zero frequencies with azimuthal
angles $\phi=0$ and $\phi=0.2\pi$, respectively, and are determined by
$\mu_r(\omega_{u/d})\cos^2\phi+\sin^2\phi=0$.
In the region between $\omega_u$ and $\omega_d$, the diagonal terms of permeability tensor in $x$-$z$ plane have
the opposite sign, that is, $\mu_{xx}>0$, $\mu_{yy}>0$ and $\mu_{zz}<0$~\cite{SM}. It is
similar to type-I hyperbolic metamaterial~\cite{ZhangMing,2020Near} without
considering the off-diagonal term in the permeability tensor.
Comparing with that of $\phi=0$ condition, i.e., $\mu_{xx}<0$, $\mu_{yy}>0$ and
$\mu_{zz}<0$, the twist-induced hyperbolic SMPs emerge at $x$-$z$ plane when $\phi=0.2 \pi$. Fig.~\ref{fig1}(d) proves the existence of such hyperbolic SMPs and also shows that it can provide more channels for radiative heat transfer. So this azimuthal-angle dependent hyperbolic mode can contribute to a enhancement of radiative heat transfer.
The coexistence of nonreciprocal and anisotropic effects in FMI is helpful for twisted and
magnetic thermal management.

{\it Twist-induced Near-field Thermal Switch.--}
To study the twist-induced thermal switch mediated by the nonreciprocal SMPs, the thermal
switch ratio $R_\kappa(\theta)$ is defined as
\begin{equation} \label{RTMR}
  R_\kappa(\theta) = \kappa(\theta) /\kappa_{\rm min}
\end{equation}
where $\kappa_{\rm min}$ is the minimal heat transfer coefficient by changing the twist
angle $\theta$.

\begin{figure}
\centering
  \includegraphics[width=\columnwidth]{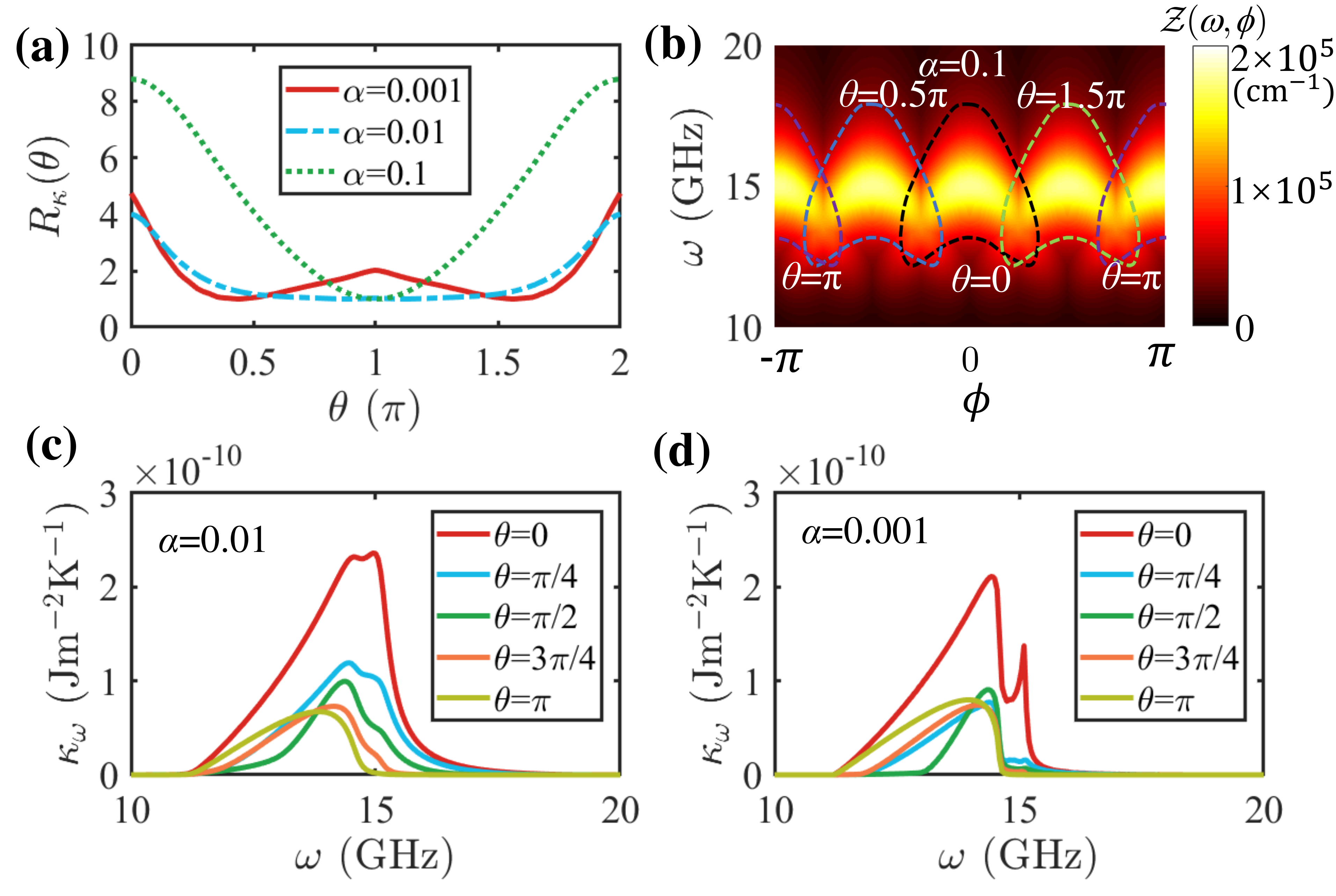} \\
  \caption{(a) Twist-induced near-field thermal switch ratio as a function of twist angle
    with different damping constants $\alpha$. (b) The contour for integrated energy
    transfer coefficient in $\omega$-$\phi$ space at single vacuum-FMI interface with
    different twist angles. (c)-(d) The spectral function of heat transfer coefficient with
    different damping constants and twist angles.}
  \label{Fig2}
\end{figure}

Fig.~\ref{Fig2}(a) shows the switch ratio with different damping constants $\alpha$. It can be seen that the switch ratio is maximal at the parallel
configuration ($\theta=0$). At large damping conditions, the green-dotted line in
Fig.~\ref{Fig2}(a) indicates that the switch ratio reaches about $9$. The physical
mechanism of such a large switch ratio can be related to the match or mismatch of the
integrated energy transmission coefficient in the $\omega$-$\phi$ space. As shown in
Fig.~\ref{Fig2}(b), the overlap region of the integrated energy transmission coefficient
reaches maximal value in parallel configuration. With increasing or decreasing the twist angle
$\theta$, the central region of the integrated energy transmission coefficient at the
twisted FMI will shift left or right in $\omega$-$\phi$ space and the overlap between two
FMIs reaches the minimum value in anti-parallel configuration. These twist-induced mismatch
effects result in a large thermal switch ratio.

\begin{figure}
\centering
  \includegraphics[width=\columnwidth]{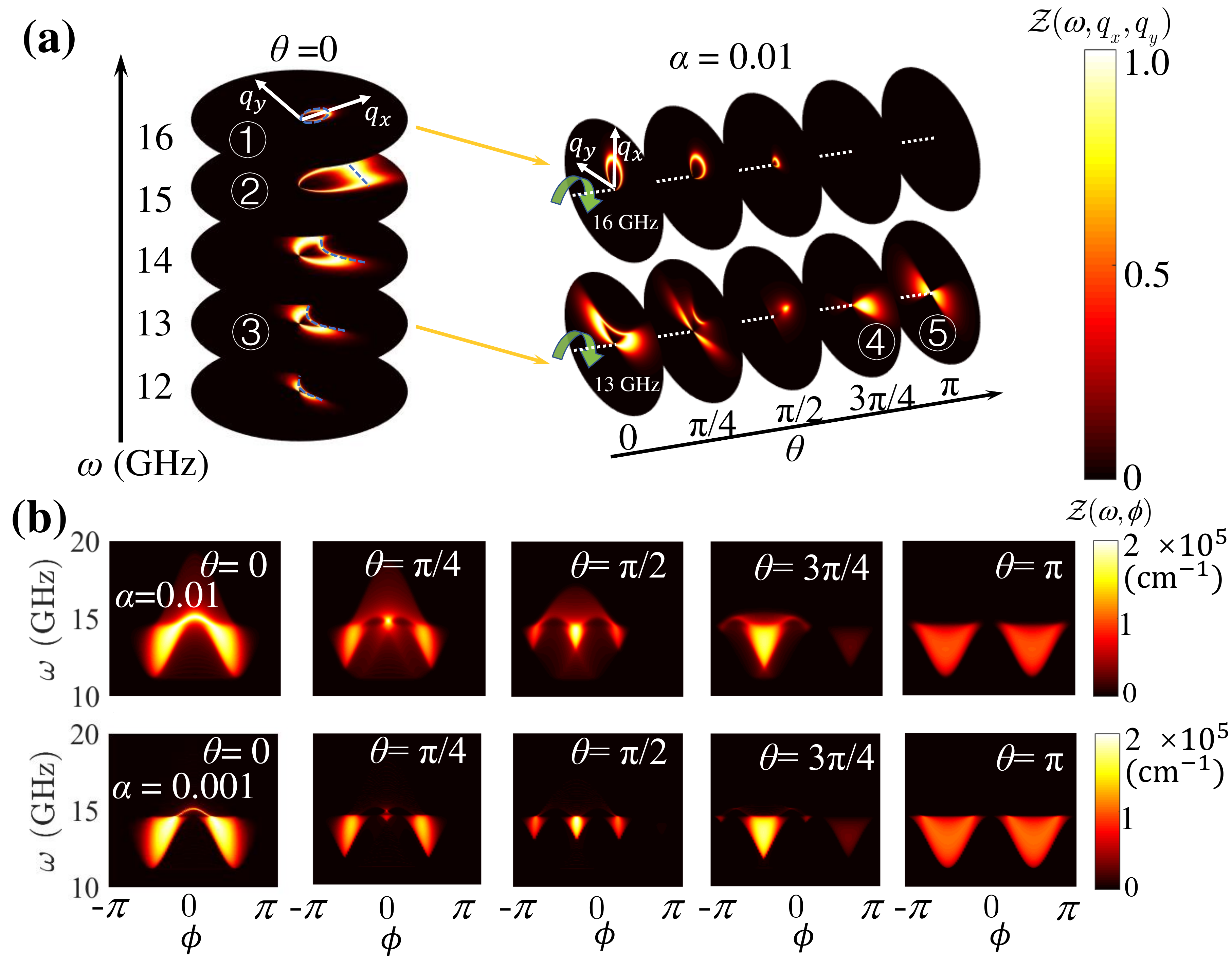} \\
  \caption{ (a) Twist-induced energy transmission coefficient with different frequency in $q_x$-$q_y$ space. Left-vertical slice figures are the energy
transmission coefficient with zero twist angle with frequency increasing. Right-transverse
slice figures are the energy transmission coefficient at fixed frequency with the twist
angle increasing. (b) Integrated energy transmission coefficient in $\omega$-$\phi$ space
with different twist angles and damping constants. }
  \label{Fig3}
\end{figure}

Under a small damping, the switch ratio is nonmonotonic with respect to the twist
angle, as indicated by the red solid line in Fig.~\ref{Fig2}(a). Such angle-dependent
behavior is similar to the thermal magnetoresistance between two
magneto-optical plasmonic particles at a large applied magnetic field~\cite{GTM2017}.
To explore this different angle dependence at small damping condition, we show the
spectral function $\kappa_{\omega}$ by varying the twist angle in Figs.~\ref{Fig2}(c) and
\ref{Fig2}(d).
The twist angle strongly modulates the height and the width of the spectral function peaks
at $0<\theta<\pi/2$. However, when $\pi/2 <\theta < \pi$, the high-frequency peak
in spectral function almost disappear and the width of the low-frequency peak becomes broader with $\theta$ increasing.
Such results qualitatively indicate that there are several nonreciprocal SMPs taking
part in the heat transfer with different angle dependence. The isofrequency contour for
energy transmission coefficient at $q_x$-$q_y$ space in Fig.~\ref{Fig3}(a) numerically
verify that statement and we show three kinds of SMPs: elliptic SMPs, hyperbolic SMPs~\cite{ZhangMing}, and twist-non-resonant SMPs. The different twist-induced tunneling and competition between
those modes lead to above nonmonotonic twist manipulation for heat transfer.

Figure~\ref{Fig3}(a) shows the different twist-induced energy transmission coefficient
of the above mentioned SMPs in $q_x$-$q_y$ space. Due to the nonreciprocal properties of
SMPs, the tunneling of three kinds of SMPs only occur at a positive $q_x$ region, except in
antiparallel configuration. On the one hand, the vertical slice contours in
Fig.~\ref{Fig3}(a) indicate that there is a transition between hyperbolic SMPs and
elliptic SMPs with an increase of frequency. We highlight that the isofrequency contours of
the energy transmission coefficient can be almost flat at $\omega \approx 15\,$GHz and
result in a sharp peak in the spectral function (Figs.~\ref{Fig2}(c) and~\ref{Fig2}(d)).
In that scenario, such flattening transition behavior allows the SMPs bands of each
individual FMI hybridize and strongly coupled to each other with large wavenumbers and
involves a dramatic increase of the local density of states for near-field radiative heat
transfer. On the other hand, Fig.~\ref{Fig3}(a) also indicates that the elliptic SMPs and
hyperbolic SMPs propagate at the open-angle ($-\phi_m<\phi<\phi_m$), where $\phi_m=\arctan
\sqrt{1/[\mu_i(\omega)-\mu_r(\omega)]}$. But the twist-non-resonant SMPs emerge when
$\omega < 15\ $GHz and is not bounded by the open-angle $\phi_m$ because it originates in
the twist-induced anisotropic in $x-z$ plane. The horizontal slice figures in
Fig.~\ref{Fig3}(a) demonstrate the twist-induced effects of three kinds of SMPs:
monotonically decreasing for elliptic SMPs and hyperbolic SMPs and nonmonotonic dependence
for twist-non-resonant SMPs at $0<\theta<\pi$. The competition mechanism among three kinds
of modes can be understood from the integrated energy transmission coefficient in
$\omega$-$\phi$ space with different damping constants (Fig.~\ref{Fig3}(b)). When $\alpha =
0.01$, elliptic SMPs, and hyperbolic SMPs play an equal role for radiative heat
transfer comparing with twist-non-resonant SMPs, which leads to an almost
monotonically decreased thermal switch ratio. In the small damping condition, i.e., $\alpha =0.001$, the twist-non-resonant SMPs will play the dominant role for radiative heat
transfer, which is induced by the optical gyrotropy and leads to a $\theta$ anisotropy in
the radiative heat transfer.

\begin{figure}
\centering
  \includegraphics[width=\columnwidth]{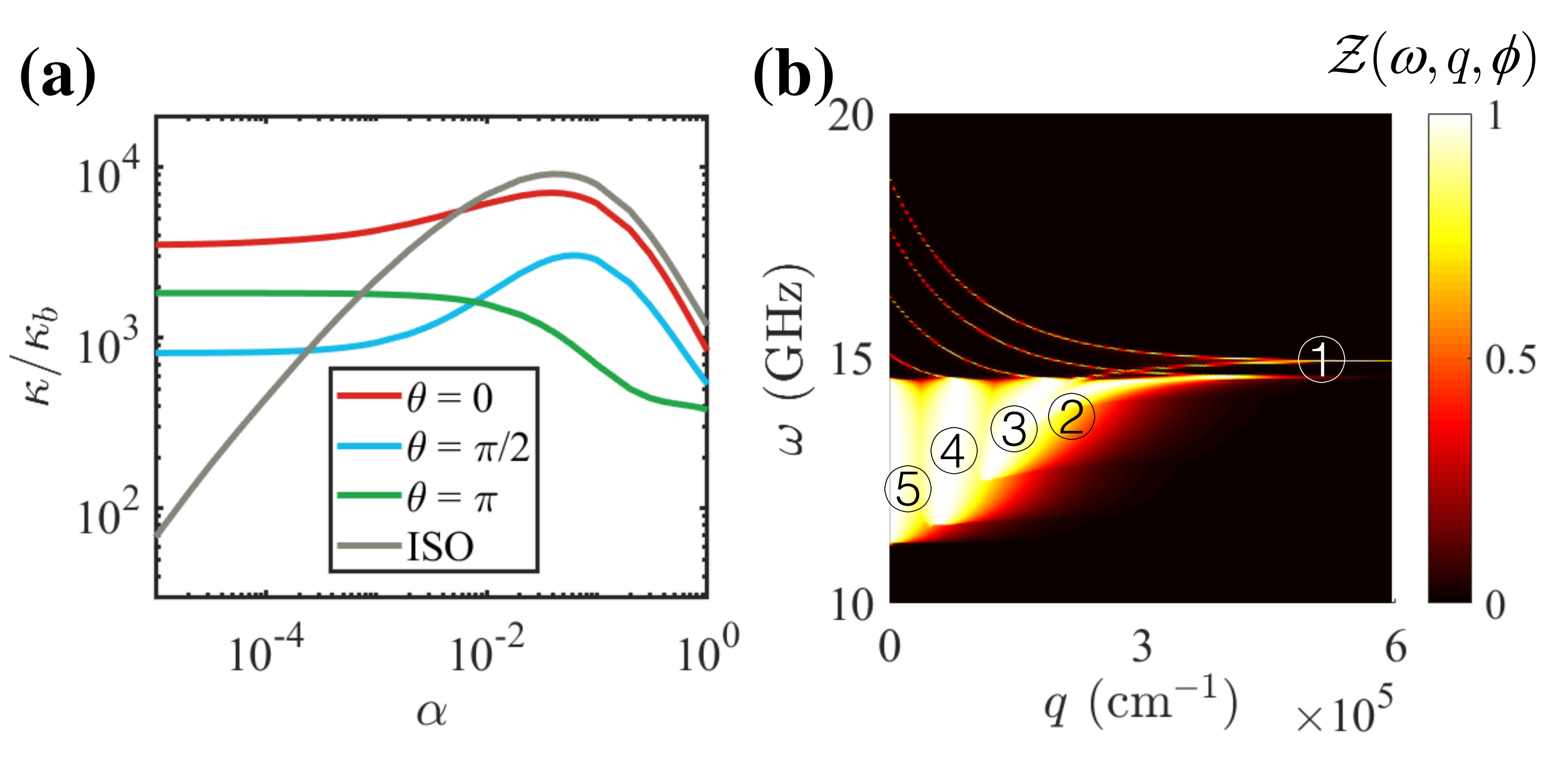} \\
  \caption{(a) Heat transfer coefficient as a function of damping constant $\alpha$ with
    different twist angle. Gray-solid line is the heat transfer coefficient between two
    isotropic slab i.e. $\mu_{xx}=\mu_{yy}=\mu_{zz}$.
    (b) Energy transmission coefficient in $\omega$-$q$ space with different azimuthal
    angle $\phi$. (\textcircled{1}-\textcircled{5}) means that the
    azimuthal angles $\phi$ are from $0.1\pi$ to $0.5\pi$ with step $0.1\pi$,
    respectively. The twist angle $\theta$ is zero and the damping constant $\alpha$ is 0.001.}
  \label{Fig4}
\end{figure}

Besides, we find an optimal damping constant for maximizing the heat transfer coefficient in Fig.~\ref{Fig4}(a): the magnitude of heat current can be enhanced almost one order in
ultra-small damping condition comparing with the isotropic case and the heat flux is
monotonically decreased at antiparallel configuration ($\theta=\pi$). Based on fluctuation
electrodynamics, heat flux between two semi-infinite systems is proportional to the
imaginary part of the permeability and the magnitude of heat current could be reduced to
zero when the damping constant approach zero or a large value. But the heat transfer
coefficient between two FMIs reaches a fixed value in zero damping constant limit. We demonstrate that twist-non-resonant SMPs play the dominant role in ultra-small damping conditions and the mechanism is slightly different from near-field radiative heat transfer in multilayer structure due to multiple surface-states coupling~\cite{Iizuka2018}. The intrinsic relation between twist-non-resonant SMPs and the heat transfer coefficient is demonstrated at Fig.~\ref{Fig4}(b): the near-unity region in energy transmission coefficient contour with different azimuthal angle can fill in the giant area in $\omega - q$ space and the local density of states for twist-non-resonant SMPs can be boosted without the constraint of ultra-small damping condition. It also demonstrates that the local density of states for elliptic SMPs and hyperbolic SMPs (the thin-solid line in Fig.~\ref{Fig4}(b)) is not be enhanced and plays little contribution for heat transfer in a ultra-small damping condition. As a whole, the different $\alpha$ and $\theta$ dependence of elliptic SMPs, hyperbolic SMPs, and twist-non-resonant SMPs result in above twist-induced manipulation for near-field radiative heat transfer.

To conclude, we have studied twist-induced near-field radiative heat transfer between two
FMIs through nonreciprocal SMPs. We find a large and nonmonotonic
twist-induced near-field thermal switch ratio. In addition, the near-field radiative heat
transfer can be enhanced by the contributions from the twist-non-resonant SMPs under ultra-small damping condition. Our results provide insights for active near-field heat transfer control by engineered twists.

\begin{acknowledgement}
J.-P., L.-W., H.-C., and J.-R. are supported by the National Key Research Program of China (Grant No. 2016YFA0301101), National Natural Science Foundation of China (No. 11935010, No. 11775159 and No. 61621001), the Shanghai Science and Technology Committee (Grants No. 18ZR1442800 and No. 18JC1410900), and the Opening Project of Shanghai Key Laboratory of Special Artificial Microstructure Materials and Technology. G.-T. thanks the financial support from the Swiss National Science Foundation (SNSF) and the NCCR Quantum Science and Technology. Rair Macedo acknowledges support from the Leverhulme Trust and the University of Glasgow through LKAS funds.
\end{acknowledgement}

\begin{suppinfo}
In the Supplemental Material, we derive the dispersion relation and the reflection
coefficients of the surface magnon polariton.

\subsection{Dispersion relation of surface magnon polariton}
To find the dispersion relation of surface magnon polaritons (SMP) at a single
vacuum-FMI interface, we employ Maxwell equations,
\begin{align}
  \nabla \times {\bf E} &= -\partial_t {\bf B} , \label{Maxwell1} \\
  \nabla \times {\bf H} &= \partial_t {\bf D} ,  \label{Maxwell2}
\end{align}
with ${\bf B}=\mu_0\mu {\bf H}$ and ${\bf D}=\epsilon_0\epsilon {\bf E}$. The SMP is
transverse electric (TE or $s$) polarized and the transverse magnetic (TM or
$p$-polarized) mode does not exist for the case where only single FMI slab is
considered.
By applying an in-plane magnetic field along the $y$-direction, the electric fields in
vacuum ${\bf E}_0$ and in FMI ${\bf E}_1$ propagate along the $x$-direction and
decay along the $z$-direction with the expressions
\begin{align}
  {\bf E}_0(x,z,t) &=\hat{y} E e^{iqx - i\beta_0 z}e^{-i\omega t},
  \quad &&{\rm Im}(\beta_0)<0, \\
  {\bf E}_1(x,z,t) &=\hat{y} E e^{iqx + i\beta_1 z}e^{-i\omega t},
  \quad &&{\rm Im}(\beta_1)<0,
\end{align}
where $q$ is the in-plane wave vector along the $x$-direction. The out-of-plane wave
vector in vacuum and FMI are denoted as $\beta_0$ and $\beta_1$, respectively. The
corresponding magnetic fields are expressed as
\begin{equation}
  {\bf B}_{0/1}(x,z,t) =\frac{i}{\omega} (\hat{x}\partial_z -\hat{z}\partial_x) ({\bf
  E}_{0/1} \cdot \hat{y}) .
\end{equation}
From ${\bf H}=(\mu_0\mu)^{-1}{\bf B}$, the magnetic field strengths are
\begin{align}
  {\bf H}_0 =& \frac{i}{\omega\mu_0} [\hat{x} \partial_z - \hat{z}\partial_x ]({\bf
  E}_0\cdot \hat{y}) , \\
  {\bf H}_1 =& \frac{i}{\omega\mu_0 (\mu_r^2-\mu_i^2)} [\hat{x}(-i\mu_i\partial_x
  +\mu_r\partial_z) + \hat{z}(-\mu_r\partial_x-i\mu_i\partial_z) ]({\bf E}_1\cdot\hat{y}) .
\end{align}
Using Eq.~\eqref{Maxwell2} in both vacuum and FMI, one has
\begin{align}
  &\beta_0^2 + q^2 = k_0^2, \label{beta0} \\
  &\beta_1^2 + q^2 = \epsilon \mu_{\rm eff} k_0^2, \label{beta1}
\end{align}
with $k_0=\omega/c$ and $\mu_{\rm eff}=(\mu_r^2-\mu_i^2)/\mu_r$.
Using the interface conditions for the magnetic field strengths, ${\bf H}_0\cdot\hat{x}
={\bf H}_1\cdot\hat{x}$, the implicit dispersion relation for the SMPs can be obtained
with
\begin{equation} \label{dispersion}
  \beta_0 + (\mu_r\beta_1 -i\mu_i q)/({\mu_r^2-\mu_i^2})=0 .
\end{equation}
From Eqs.~\eqref{beta0}, \eqref{beta1} and \eqref{dispersion}, the dispersion relation of
SMP can be numerically obtained.

\subsection{Reflection coefficients}
In this section, we obtain the reflection coefficients by taking the anisotropic effect
into account. For the case where incidence plane is at an angle $\phi$ with respect to
the $x$-axis, the effective permeability tensor is
\begin{equation}
  \mu'= {\cal R} \mu {\cal R}^T =
  \begin{bmatrix}
    \mu_{xx} & \mu_{xy} & \mu_{xz} \\
    \mu_{yx} & \mu_{yy} & \mu_{yz} \\
    \mu_{zx} & \mu_{zy} & \mu_{zz}
  \end{bmatrix} ,
\end{equation}
where ${\cal R}$ is the rotation matrix with
\begin{equation} \label{phi_rota}
  {\cal R} =
  \begin{bmatrix}
    \cos\phi & \sin\phi & 0 \\  -\sin\phi & \cos\phi & 0 \\ 0 & 0 & 1
  \end{bmatrix} .
\end{equation}
Although the SMP is $s$-polarized by considering a single FMI slab, the
$p$-polarized mode exists between two FMI slabs as well due to the anisotropic
permeability tensor.
We focus on the interface between the lower FMI slab and vacuum. The genernal
form of the electric and magnetic fields inside the FMI slab can be written as
\begin{align}
  {\bf E} &= ({\cal E}_x, {\cal E}_y, {\cal E}_z) e^{-i\omega t +iqx}, \\
  {\bf H} &= ({\cal H}_x, {\cal H}_y, {\cal H}_z) e^{-i\omega t +iqx},
\end{align}
where the superscript $'$ in the space variables $x'$, $y'$ and $z'$ is dropped for
simplicity. From the Maxwell equations, Eqs.~\eqref{Maxwell1} and \eqref{Maxwell2}, we can
get the differential equation
\begin{equation}
  \frac{d}{dz}
  \begin{bmatrix}
    {\cal E}_x \\ {\cal E}_y \\ \alpha{\cal H}_x \\ \alpha{\cal H}_y
  \end{bmatrix}
  = i K
  \begin{bmatrix}
    {\cal E}_x \\ {\cal E}_y  \\ \alpha{\cal H}_x \\ \alpha{\cal H}_y
  \end{bmatrix}
\end{equation}
with $\alpha=\sqrt{\mu_0/\epsilon_0}$ and
\begin{equation}
  K =
  \begin{bmatrix}
    0 & q\mu_{yz}/\mu_{zz} & k_0(\mu_{yx}-\mu_{yz}\mu_{zx}/\mu_{zz}) &
    k_0(\mu_{yy}-\mu_{yz}\mu_{zy}/\mu_{zz})- q^2/(k_0\epsilon) \\
    0 & -q\mu_{xz}/\mu_{zz} & k_0(-\mu_{xx}+\mu_{xz}\mu_{zx}/\mu_{zz}) &
    k_0(-\mu_{xy}+\mu_{xz}\mu_{zy}/\mu_{zz}) \\
    0 & -k_0\epsilon+q^2/(k_0\mu_{zz}) & -q\mu_{zx}/\mu_{zz} & -q\mu_{zy}/\mu_{zz} \\
    k_0\epsilon & 0 & 0 & 0
  \end{bmatrix}.
\end{equation}

By solving this differential equation, we get
\begin{align}
   [{\cal E}_x(z), {\cal E}_y(z), \alpha{\cal H}_y(z), \alpha{\cal H}_y(z)]= \sum\nolimits_{m=1}^{2}c_m[u_{1,m}, u_{2,m}, u_{3,m}, u_{4,m}] e^{ik_m z},
\end{align}
where $k_m$ and $u_{i,m}$ are, respectively, the eigenvalue and eigenvector of matrix $K$.
Since $K$ is a four-by-four matrix, we have four eigenvalues: two of them satisfy ${\rm
Im}(k_m)<0$ and the other two ${\rm Im}(k_m)>0$. We take $k_m$ with ${\rm Im}(k_m)<0$, of
which the subscripts are denoted as $m=1, 2$, to ensure that the electromagnetic fields
vanish at $z{\rightarrow}{-\infty}$.

In the vacuum, the incoming electric and magnetic fields can be, respectively, writen as
\begin{align}
  {\bf E}_{\rm in} =& \left[ e_{\rm in}^s \hat{y} +e_{\rm in}^p (\beta_0 \hat{x} -q
  \hat{z})/k_0 \right] e^{i\omega t -iqx -i\beta_0 z} , \\
  \alpha{\bf H}_{\rm in} =& \left[ e_{\rm in}^p \hat{y} -e_{\rm in}^s (\beta_0 \hat{x} -q
  \hat{z})/k_0 \right] e^{i\omega t -iqx -i\beta_0 z} ,
\end{align}
where the superscripts $s$ and $p$ are used to denote the polarizations. The reflected
fields are expressed as
\begin{align}
  {\bf E}_{\rm re} =& \left[ e_{\rm re}^s \hat{y} -e_{\rm re}^p (\beta_0 \hat{x} +q
  \hat{z})/k_0 \right] e^{i\omega t -iqx +i\beta_0 z}, \\
  \alpha{\bf H}_{\rm re} =& \left[ e_{\rm re}^p \hat{y} +e_{\rm re}^s (\beta_0 \hat{x} +q
  \hat{z})/k_0 \right] e^{i\omega t -iqx +i\beta_0 z}.
\end{align}
At the interface of the vacuum side with $z=0^+$, the in-plane electric and magnetic fields
can be written as
\begin{align}
  {\bf E}_{\parallel} =& [ (e_{\rm in}^s +e_{\rm re}^s) \hat{y} + (e_{\rm in}^p -e_{\rm
  re}^p) \beta_0/k_0 \hat{x} ] e^{i\omega t -iqx} , \\
  \alpha{\bf H}_{\parallel} =& [ (e_{\rm in}^p +e_{\rm re}^p) \hat{y} - (e_{\rm in}^s
  -e_{\rm re}^s) \beta_0/k_0 \hat{x} ] e^{i\omega t -iqx} .
\end{align}

For the case of the $s$-polarized incoming field, that is, $e_{\rm in}^p =0$, the
interface conditions give
\begin{align}
  -e_{\rm re}^p \beta_0/k_0 =& \ {\cal E}_x(z=0) ,\label{es1} \\
  e_{\rm in}^s +e_{\rm re}^s =& \ {\cal E}_y(z=0) ,\label{es2} \\
  -(e_{\rm in}^s -e_{\rm re}^s) \beta_0/k_0 =& \ \alpha{\cal H}_x(z=0) ,\label{es3} \\
  e_{\rm re}^p =& \ \alpha{\cal H}_y(z=0) .\label{es4}
\end{align}
From Eqs.~\eqref{es1} and \eqref{es4}, we have
\begin{equation} \label{c2c1_s}
  c_2/c_1=-(u_{1,1}k_0 +u_{4,1}\beta_0)/(u_{1,2}k_0 +u_{4,2}\beta_0).
\end{equation}
The reflection coefficient $r_{ss}=e_{\rm re}^s/e_{\rm in}^s$ can be obtained from
Eqs.~\eqref{es2} and \eqref{es3} as
\begin{equation}
  r_{ss} =\frac{(u_{2,1}\beta_0 +u_{3,1}k_0)+(u_{2,2}\beta_0 +u_{3,2}k_0)
  c_2/c_1}{(u_{2,1}\beta_0 -u_{3,1}k_0)+(u_{2,2}\beta_0 -u_{3,2}k_0)c_2/c_1}.
\end{equation}
From Eqs.~\eqref{es1} and \eqref{es3}, we can obtain $r_{ps}=e_{\rm re}^p/e_{\rm in}^s$ as
\begin{equation}
  r_{ps} =(1-r_{ss})\frac{u_{1,1}+u_{1,2}c_2/c_1}{u_{3,1}+u_{3,2}c_2/c_1}.
\end{equation}

For the case of the $p$-polarized incoming field, that is, $e_{\rm in}^s =0$, the
interface conditions give
\begin{align}
  (e_{\rm in}^p -e_{\rm re}^p) \beta_0/k_0 =& \ {\cal E}_x(z=0) ,\label{ep1} \\
  e_{\rm re}^s =& \ {\cal E}_y(z=0)  ,\label{ep2} \\
  e_{\rm re}^s \beta_0/k_0 =& \ \alpha{\cal H}_x(z=0) ,\label{ep3} \\
  e_{\rm in}^p +e_{\rm re}^p =& \ \alpha{\cal H}_y(z=0) .\label{ep4}
\end{align}
From Eqs.~\eqref{ep2} and \eqref{ep3}, we have
\begin{equation} \label{c2c1_p}
  d_2/d_1 \equiv c_2/c_1=-(u_{2,1}\beta_0 -u_{3,1}k_0)/(u_{2,2}\beta_0 -u_{3,2}k_0).
\end{equation}
Notice $c_2/c_1$ here is different from that in Eq.~\eqref{c2c1_s} and we denote it as
$d_2/d_1$ instead.
The reflection coefficient $r_{pp}=e_{\rm re}^p/e_{\rm in}^p$ obtained from
Eqs.~\eqref{ep2} and \eqref{ep3} is expressed as
\begin{equation}
  r_{pp} =\frac{(u_{4,1}\beta_0 -u_{1,1}k_0)+(u_{4,2}\beta_0 -u_{1,2}k_0)
  d_2/d_1}{(u_{4,1}\beta_0 +u_{1,1}k_0)+(u_{4,2}\beta_0 +u_{1,2}k_0)d_2/d_1}.
\end{equation}
From Eqs.~\eqref{ep1} and \eqref{ep3}, we obtain $r_{sp}=e_{\rm re}^s/e_{\rm in}^p$ as
\begin{equation}
  r_{sp} =(1-r_{pp})\frac{u_{3,1}+u_{3,2}d_2/d_1}{u_{1,1}+u_{1,2}d_2/d_1}.
\end{equation}

\subsection{Near-field radiative heat transfer}
From the fluctuating electrodynamics, the radiative heat current with vacuum gap $d$ is given by
\begin{equation}
  J =\int_{0}^{\infty} \frac{d\omega}{2\pi} \hbar\omega(N_1-N_2) \int_{0}^{\infty}
  \frac{dq}{2\pi} q \int_{0}^{2\pi} \frac{d\phi}{2\pi} {\cal Z}(\omega,q,\phi),
\end{equation}
where $N_{i}=1/[e^{\hbar\omega/(k_B T_i)}-1]$ with $i=1,2$ is the Bose-Einstein
distribution function. The photonic transmission coefficient ${\cal Z}(\omega,q,\phi)$ reads
\begin{equation}
  {\cal Z} =
  \begin{cases}
    {\rm Tr}[({\bf I}-{\bf R}_2^*{\bf R}_2){\bf D}({\bf I}-{\bf R}_1^*{\bf R}_1){\bf
    D}^*], & q<\omega/c \\
    {\rm Tr}[({\bf R}_2^*-{\bf R}_2){\bf D}({\bf R}_1^*-{\bf R}_1){\bf
    D}^*]e^{-2|\beta_0|d}, & q>\omega/c
  \end{cases}
\end{equation}
where $q$ and $\beta_0=\sqrt{(\omega/c)^2-q^2}$ are the in-plane and out-of-plane wave
vectors, respectively. The identity matrix is denoted as ${\bf I}$.
The reflection coefficient matrix for the interface between vacuum and the FMI
$i$ is
\begin{equation}
  {\bf R}_i=
  \begin{bmatrix}
    r_{ss}^i & r_{sp}^i \\ r_{ps}^i & r_{pp}^i
  \end{bmatrix} .
\end{equation}
The Fabry-Perot-like matrix reads ${\bf D}=({\bf I}-{\bf R}_1{\bf R}_2 e^{2i\beta_0
d})^{-1}$.

The heat transfer coefficient, which is defined as $\kappa(T) \equiv \lim_{\Delta T
\rightarrow 0} J / \Delta T$, is expressed as
\begin{align}
  \kappa(T) =\int_0^{\infty} \frac{d\omega}{2\pi} \hbar\omega N' \int_{0}^{\infty}
  \frac{dq}{2\pi} q \int_{0}^{2\pi} \frac{d\phi}{2\pi} Z(\omega,q,\phi),
\end{align}
where the derivative of the Bose-Einstein distribution with respect to the temperature is
expressed as
\begin{equation}
  N' \equiv \partial N/\partial T =\frac{\hbar\omega\, e^{\hbar\omega/(k_BT)}}{k_B T^2
  \left[ e^{\hbar\omega/(k_BT)} -1 \right]^2}.
\end{equation}

For the case of $\phi=0$, the reflection coefficient for the $s$-polarized mode can be
expressed as
\begin{equation}
  r_{ss} =\frac{\beta_0-(\beta_1 \mu_r -iq \mu_i) /({\mu_r^2-\mu_i^2})}
  {\beta_0+(\beta_1 \mu_r - iq \mu_i)/({\mu_r^2-\mu_i^2})},
\end{equation}
and the other reflection coefficients vanish. The photonic transmission coefficient can be
expressed as,
\begin{align}
  {\cal Z}(\omega,q,\phi=0) = \frac{4[{\rm Im}(r_{ss})]^2 e^{-2|\beta_0|d}}{|1-r_{ss}^2
  e^{-2|\beta_0|d}|^2} ,
\end{align}
from which one can obtain the nonreciprocal dispersion of the symmetric and asymmetric SMP
modes between two FMI slabs.
\subsection{Permeability tensor components of uniaxial FMI}
\begin{figure}
\centering
  \includegraphics[width=\columnwidth]{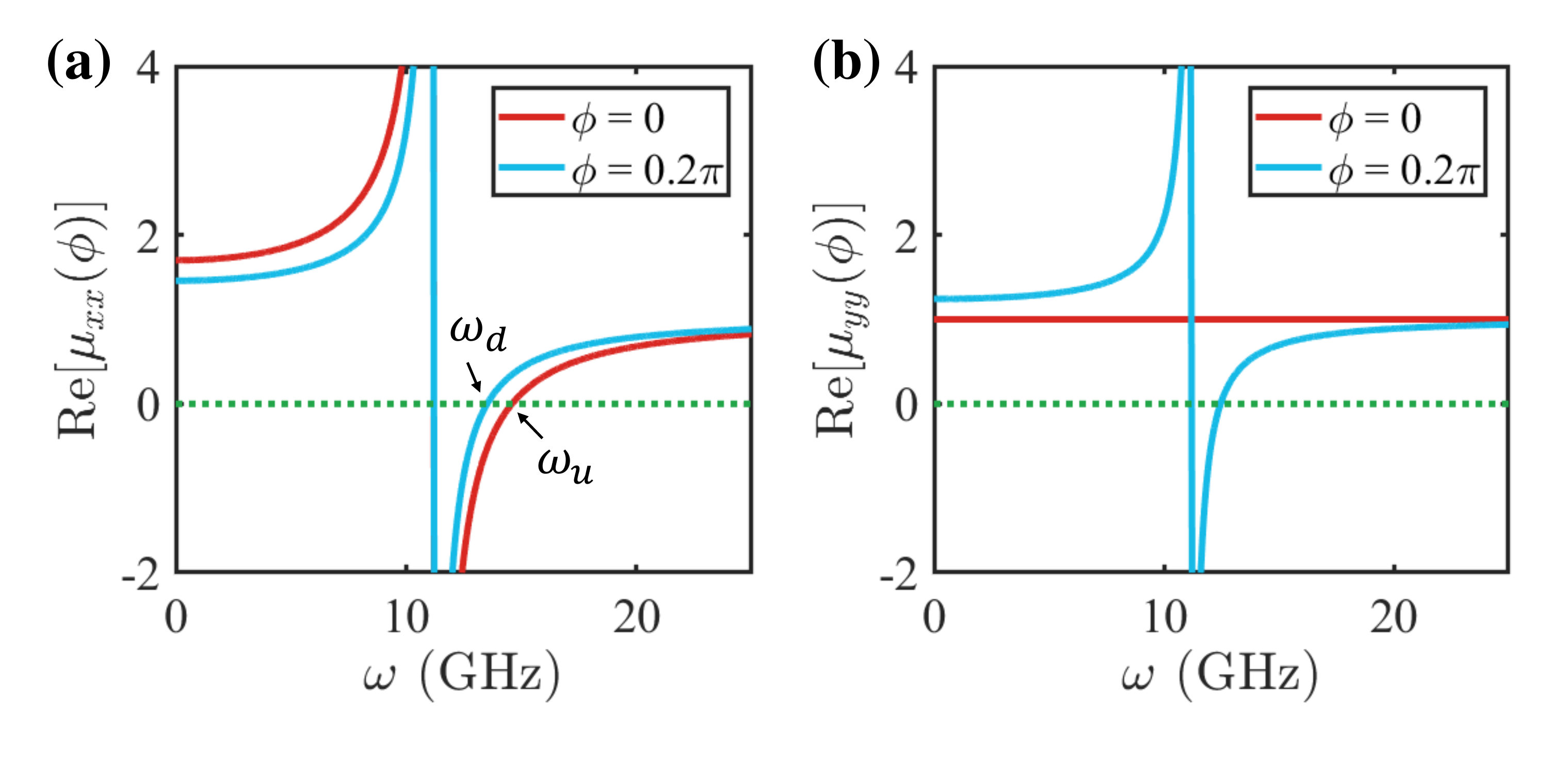} \\
  \caption{(a-b) The real part of diagonal term of permeability tensor with different azimuthal angle. The damping factor is $\alpha$=0.01.}
  \label{FigS1}
\end{figure}
In our calculation, we use the parameters of YIG and rewrite the permeability tensor as below:
\begin{equation}
  \mu =
  \begin{bmatrix}
  \mu_r & 0 & -i\mu_i \\  0 & 1 & 0 \\ i\mu_i & 0 & \mu_r
  \end{bmatrix}
\end{equation}
After a $\phi$ rotation in $x-y$ plane, the rotated permeability tensor regards as:
\begin{equation}
  \mu(\phi) =
  \begin{bmatrix}
  \mu_r \cos(\phi)^2+\sin(\phi)^2 & \cos(\phi)\sin(\phi)(\mu_r-1)  & -i\mu_i \cos(\phi) \\
   \cos(\phi)\sin(\phi)(\mu_r-1) &  \cos(\phi)^2+\mu_r\sin(\phi)^2  & -i\mu_i \sin(\phi) \\
   i\mu_i \cos(\phi) & i\mu_i \sin(\phi) & \mu_r
  \end{bmatrix}
\end{equation}
Fig.~\ref{FigS1} shows the real part of $\mu_{xx}$ and $\mu_{yy}$ with different azimuthal angle and $\mu_zz$ = $\mu_r$=$\mu_{xx}(\phi)=0$. It indicates value of $\omega_u$ and $\omega_d$ presented in the main page at the $\mu$-near-zero frequency.
\end{suppinfo}

\bibliography{bib_near-field}{}

\providecommand{\latin}[1]{#1}
\makeatletter
\providecommand{\doi}
  {\begingroup\let\do\@makeother\dospecials
  \catcode`\{=1 \catcode`\}=2 \doi@aux}
\providecommand{\doi@aux}[1]{\endgroup\texttt{#1}}
\makeatother
\providecommand*\mcitethebibliography{\thebibliography}
\csname @ifundefined\endcsname{endmcitethebibliography}
  {\let\endmcitethebibliography\endthebibliography}{}
\begin{mcitethebibliography}{56}
\providecommand*\natexlab[1]{#1}
\providecommand*\mciteSetBstSublistMode[1]{}
\providecommand*\mciteSetBstMaxWidthForm[2]{}
\providecommand*\mciteBstWouldAddEndPuncttrue
  {\def\EndOfBibitem{\unskip.}}
\providecommand*\mciteBstWouldAddEndPunctfalse
  {\let\EndOfBibitem\relax}
\providecommand*\mciteSetBstMidEndSepPunct[3]{}
\providecommand*\mciteSetBstSublistLabelBeginEnd[3]{}
\providecommand*\EndOfBibitem{}
\mciteSetBstSublistMode{f}
\mciteSetBstMaxWidthForm{subitem}{(\alph{mcitesubitemcount})}
\mciteSetBstSublistLabelBeginEnd
  {\mcitemaxwidthsubitemform\space}
  {\relax}
  {\relax}

\bibitem[Planck and Masius(1914)Planck, and Masius]{Planck}
Planck,~M.; Masius,~M. \emph{The Theory of Heat Radiation}; P. Blakiston’s
  Son \& Co, New York, 1914\relax
\mciteBstWouldAddEndPuncttrue
\mciteSetBstMidEndSepPunct{\mcitedefaultmidpunct}
{\mcitedefaultendpunct}{\mcitedefaultseppunct}\relax
\EndOfBibitem
\bibitem[Cahill \latin{et~al.}(2014)Cahill, Braun, Chen, Clarke, Fan, Goodson,
  Keblinski, King, Mahan, Majumdar, Maris, Phillpot, Pop, and Shi]{thermal-APR}
Cahill,~D.~G.; Braun,~P.~V.; Chen,~G.; Clarke,~D.~R.; Fan,~S.; Goodson,~K.~E.;
  Keblinski,~P.; King,~W.~P.; Mahan,~G.~D.; Majumdar,~A.; Maris,~H.~J.;
  Phillpot,~S.~R.; Pop,~E.; Shi,~L. Nanoscale thermal transport. II.
  2003–2012. \emph{Appl. Phys. Rev.} \textbf{2014}, \emph{1}, 011305\relax
\mciteBstWouldAddEndPuncttrue
\mciteSetBstMidEndSepPunct{\mcitedefaultmidpunct}
{\mcitedefaultendpunct}{\mcitedefaultseppunct}\relax
\EndOfBibitem
\bibitem[Polder and Van~Hove(1971)Polder, and Van~Hove]{PvH}
Polder,~D.; Van~Hove,~M. Theory of Radiative Heat Transfer between Closely
  Spaced Bodies. \emph{Phys. Rev. B} \textbf{1971}, \emph{4}, 3303--3314\relax
\mciteBstWouldAddEndPuncttrue
\mciteSetBstMidEndSepPunct{\mcitedefaultmidpunct}
{\mcitedefaultendpunct}{\mcitedefaultseppunct}\relax
\EndOfBibitem
\bibitem[Volokitin and Persson(2007)Volokitin, and Persson]{review07}
Volokitin,~A.~I.; Persson,~B. N.~J. Near-field radiative heat transfer and
  noncontact friction. \emph{Rev. Mod. Phys.} \textbf{2007}, \emph{79},
  1291--1329\relax
\mciteBstWouldAddEndPuncttrue
\mciteSetBstMidEndSepPunct{\mcitedefaultmidpunct}
{\mcitedefaultendpunct}{\mcitedefaultseppunct}\relax
\EndOfBibitem
\bibitem[Song \latin{et~al.}(2015)Song, Fiorino, Meyhofer, and Reddy]{review15}
Song,~B.; Fiorino,~A.; Meyhofer,~E.; Reddy,~P. Near-field radiative thermal
  transport: From theory to experiment. \emph{AIP Adv.} \textbf{2015},
  \emph{5}, 053503\relax
\mciteBstWouldAddEndPuncttrue
\mciteSetBstMidEndSepPunct{\mcitedefaultmidpunct}
{\mcitedefaultendpunct}{\mcitedefaultseppunct}\relax
\EndOfBibitem
\bibitem[Cuevas and García-Vidal(2018)Cuevas, and García-Vidal]{review18}
Cuevas,~J.~C.; García-Vidal,~F.~J. Radiative Heat Transfer. \emph{ACS
  Photonics} \textbf{2018}, \emph{5}, 3896--3915\relax
\mciteBstWouldAddEndPuncttrue
\mciteSetBstMidEndSepPunct{\mcitedefaultmidpunct}
{\mcitedefaultendpunct}{\mcitedefaultseppunct}\relax
\EndOfBibitem
\bibitem[Liu \latin{et~al.}(2015)Liu, Wang, and Zhang]{review15-2}
Liu,~X.; Wang,~L.; Zhang,~Z.~M. Near-Field Thermal Radiation: Recent Progress
  and Outlook. \emph{Nanoscale and Microscale Thermophys. Eng.} \textbf{2015},
  \emph{19}, 98--126\relax
\mciteBstWouldAddEndPuncttrue
\mciteSetBstMidEndSepPunct{\mcitedefaultmidpunct}
{\mcitedefaultendpunct}{\mcitedefaultseppunct}\relax
\EndOfBibitem
\bibitem[Kim \latin{et~al.}(2015)Kim, Song, Fern{\'a}ndez-Hurtado, Lee, Jeong,
  Cui, Thompson, Feist, Reid, Garc{\'i}a-Vidal, Cuevas, Meyhofer, and
  Reddy]{Kim2015}
Kim,~K.; Song,~B.; Fern{\'a}ndez-Hurtado,~V.; Lee,~W.; Jeong,~W.; Cui,~L.;
  Thompson,~D.; Feist,~J.; Reid,~M. T.~H.; Garc{\'i}a-Vidal,~F.~J.;
  Cuevas,~J.~C.; Meyhofer,~E.; Reddy,~P. Radiative heat transfer in the extreme
  near field. \emph{Nature} \textbf{2015}, \emph{528}, 387\relax
\mciteBstWouldAddEndPuncttrue
\mciteSetBstMidEndSepPunct{\mcitedefaultmidpunct}
{\mcitedefaultendpunct}{\mcitedefaultseppunct}\relax
\EndOfBibitem
\bibitem[Cui \latin{et~al.}(2017)Cui, Jeong, Fern{\'a}ndez-Hurtado, Feist,
  Garc{\'i}a-Vidal, Cuevas, Meyhofer, and Reddy]{Cui2017}
Cui,~L.; Jeong,~W.; Fern{\'a}ndez-Hurtado,~V.; Feist,~J.;
  Garc{\'i}a-Vidal,~F.~J.; Cuevas,~J.~C.; Meyhofer,~E.; Reddy,~P. Study of
  radiative heat transfer in {\AA}ngstr{\"o}m- and nanometre-sized gaps.
  \emph{Nat. Commun.} \textbf{2017}, \emph{8}, 14479\relax
\mciteBstWouldAddEndPuncttrue
\mciteSetBstMidEndSepPunct{\mcitedefaultmidpunct}
{\mcitedefaultendpunct}{\mcitedefaultseppunct}\relax
\EndOfBibitem
\bibitem[Wang and Peng(2017)Wang, and Peng]{JS1}
Wang,~J.-S.; Peng,~J. Capacitor physics in ultra-near-field heat transfer.
  \emph{Europhys. Lett.} \textbf{2017}, \emph{118}, 24001\relax
\mciteBstWouldAddEndPuncttrue
\mciteSetBstMidEndSepPunct{\mcitedefaultmidpunct}
{\mcitedefaultendpunct}{\mcitedefaultseppunct}\relax
\EndOfBibitem
\bibitem[Tang and Wang(2018)Tang, and Wang]{gm1}
Tang,~G.; Wang,~J.-S. Heat transfer statistics in extreme-near-field radiation.
  \emph{Phys. Rev. B} \textbf{2018}, \emph{98}, 125401\relax
\mciteBstWouldAddEndPuncttrue
\mciteSetBstMidEndSepPunct{\mcitedefaultmidpunct}
{\mcitedefaultendpunct}{\mcitedefaultseppunct}\relax
\EndOfBibitem
\bibitem[Tang \latin{et~al.}(2019)Tang, Yap, Ren, and Wang]{gm2}
Tang,~G.; Yap,~H.~H.; Ren,~J.; Wang,~J.-S. Anomalous Near-Field Heat Transfer
  in Carbon-Based Nanostructures with Edge States. \emph{Phys. Rev. Applied}
  \textbf{2019}, \emph{11}, 031004\relax
\mciteBstWouldAddEndPuncttrue
\mciteSetBstMidEndSepPunct{\mcitedefaultmidpunct}
{\mcitedefaultendpunct}{\mcitedefaultseppunct}\relax
\EndOfBibitem
\bibitem[Volokitin and Persson(2004)Volokitin, and Persson]{SPP1}
Volokitin,~A.~I.; Persson,~B. N.~J. Resonant photon tunneling enhancement of
  the radiative heat transfer. \emph{Phys. Rev. B} \textbf{2004}, \emph{69},
  045417\relax
\mciteBstWouldAddEndPuncttrue
\mciteSetBstMidEndSepPunct{\mcitedefaultmidpunct}
{\mcitedefaultendpunct}{\mcitedefaultseppunct}\relax
\EndOfBibitem
\bibitem[Iizuka and Fan(2015)Iizuka, and Fan]{SPP2}
Iizuka,~H.; Fan,~S. Analytical treatment of near-field electromagnetic heat
  transfer at the nanoscale. \emph{Phys. Rev. B} \textbf{2015}, \emph{92},
  144307\relax
\mciteBstWouldAddEndPuncttrue
\mciteSetBstMidEndSepPunct{\mcitedefaultmidpunct}
{\mcitedefaultendpunct}{\mcitedefaultseppunct}\relax
\EndOfBibitem
\bibitem[Boriskina \latin{et~al.}(2015)Boriskina, Tong, Huang, Zhou, Chiloyan,
  and Chen]{SPP3}
Boriskina,~S.~V.; Tong,~J.~K.; Huang,~Y.; Zhou,~J.; Chiloyan,~V.; Chen,~G.
  Enhancement and Tunability of Near-Field Radiative Heat Transfer Mediated by
  Surface Plasmon Polaritons in Thin Plasmonic Films. \emph{Photonics}
  \textbf{2015}, \emph{2}, 659--683\relax
\mciteBstWouldAddEndPuncttrue
\mciteSetBstMidEndSepPunct{\mcitedefaultmidpunct}
{\mcitedefaultendpunct}{\mcitedefaultseppunct}\relax
\EndOfBibitem
\bibitem[Ilic \latin{et~al.}(2012)Ilic, Jablan, Joannopoulos, Celanovic,
  Buljan, and Solja\ifmmode \check{c}\else \v{c}\fi{}i\ifmmode~\acute{c}\else
  \'{c}\fi{}]{SPP-graphene1}
Ilic,~O.; Jablan,~M.; Joannopoulos,~J.~D.; Celanovic,~I.; Buljan,~H.;
  Solja\ifmmode \check{c}\else \v{c}\fi{}i\ifmmode~\acute{c}\else
  \'{c}\fi{},~M. Near-field thermal radiation transfer controlled by plasmons
  in graphene. \emph{Phys. Rev. B} \textbf{2012}, \emph{85}, 155422\relax
\mciteBstWouldAddEndPuncttrue
\mciteSetBstMidEndSepPunct{\mcitedefaultmidpunct}
{\mcitedefaultendpunct}{\mcitedefaultseppunct}\relax
\EndOfBibitem
\bibitem[Ramirez \latin{et~al.}(2017)Ramirez, Shen, and
  McGaughey]{SPP-graphene2}
Ramirez,~F.~V.; Shen,~S.; McGaughey,~A. J.~H. Near-field radiative heat
  transfer in graphene plasmonic nanodisk dimers. \emph{Phys. Rev. B}
  \textbf{2017}, \emph{96}, 165427\relax
\mciteBstWouldAddEndPuncttrue
\mciteSetBstMidEndSepPunct{\mcitedefaultmidpunct}
{\mcitedefaultendpunct}{\mcitedefaultseppunct}\relax
\EndOfBibitem
\bibitem[Yu \latin{et~al.}(2017)Yu, Manjavacas, and Garc{\'i}a~de
  Abajo]{SPP-graphene3}
Yu,~R.; Manjavacas,~A.; Garc{\'i}a~de Abajo,~F.~J. Ultrafast radiative heat
  transfer. \emph{Nat. Commun.} \textbf{2017}, \emph{8}, 2\relax
\mciteBstWouldAddEndPuncttrue
\mciteSetBstMidEndSepPunct{\mcitedefaultmidpunct}
{\mcitedefaultendpunct}{\mcitedefaultseppunct}\relax
\EndOfBibitem
\bibitem[Peng and Wang()Peng, and Wang]{SPP-graphene4}
Peng,~J.; Wang,~J.-S. Current-Induced Heat Transfer in Double-Layer
  Graphene\relax
\mciteBstWouldAddEndPuncttrue
\mciteSetBstMidEndSepPunct{\mcitedefaultmidpunct}
{\mcitedefaultendpunct}{\mcitedefaultseppunct}\relax
\EndOfBibitem
\bibitem[Zhang \latin{et~al.}(2018)Zhang, Yi, and Tan]{SPP-BP}
Zhang,~Y.; Yi,~H.-L.; Tan,~H.-P. Near-Field Radiative Heat Transfer between
  Black Phosphorus Sheets via Anisotropic Surface Plasmon Polaritons. \emph{ACS
  Photonics} \textbf{2018}, \emph{5}, 3739--3747\relax
\mciteBstWouldAddEndPuncttrue
\mciteSetBstMidEndSepPunct{\mcitedefaultmidpunct}
{\mcitedefaultendpunct}{\mcitedefaultseppunct}\relax
\EndOfBibitem
\bibitem[Rousseau \latin{et~al.}(2009)Rousseau, Laroche, and Greffet]{SPP-Si1}
Rousseau,~E.; Laroche,~M.; Greffet,~J.-J. Radiative heat transfer at nanoscale
  mediated by surface plasmons for highly doped silicon. \emph{Appl. Phys.
  Lett.} \textbf{2009}, \emph{95}, 231913\relax
\mciteBstWouldAddEndPuncttrue
\mciteSetBstMidEndSepPunct{\mcitedefaultmidpunct}
{\mcitedefaultendpunct}{\mcitedefaultseppunct}\relax
\EndOfBibitem
\bibitem[Fern\'andez-Hurtado \latin{et~al.}(2017)Fern\'andez-Hurtado,
  Garc\'{\i}a-Vidal, Fan, and Cuevas]{SPP-Si2}
Fern\'andez-Hurtado,~V.; Garc\'{\i}a-Vidal,~F.~J.; Fan,~S.; Cuevas,~J.~C.
  Enhancing Near-Field Radiative Heat Transfer with {S}i-based Metasurfaces.
  \emph{Phys. Rev. Lett.} \textbf{2017}, \emph{118}, 203901\relax
\mciteBstWouldAddEndPuncttrue
\mciteSetBstMidEndSepPunct{\mcitedefaultmidpunct}
{\mcitedefaultendpunct}{\mcitedefaultseppunct}\relax
\EndOfBibitem
\bibitem[DeSutter \latin{et~al.}(2019)DeSutter, Tang, and Francoeur]{SPP-Si19}
DeSutter,~J.; Tang,~L.; Francoeur,~M. A near-field radiative heat transfer
  device. \emph{Nat. Nanotechnol.} \textbf{2019}, \emph{14}, 751--755\relax
\mciteBstWouldAddEndPuncttrue
\mciteSetBstMidEndSepPunct{\mcitedefaultmidpunct}
{\mcitedefaultendpunct}{\mcitedefaultseppunct}\relax
\EndOfBibitem
\bibitem[Ito \latin{et~al.}(2014)Ito, Matsui, and Iizuka]{SPhP1}
Ito,~K.; Matsui,~T.; Iizuka,~H. Thermal emission control by evanescent wave
  coupling between guided mode of resonant grating and surface phonon polariton
  on silicon carbide plate. \emph{Appl. Phys. Lett.} \textbf{2014}, \emph{104},
  051127\relax
\mciteBstWouldAddEndPuncttrue
\mciteSetBstMidEndSepPunct{\mcitedefaultmidpunct}
{\mcitedefaultendpunct}{\mcitedefaultseppunct}\relax
\EndOfBibitem
\bibitem[Shen \latin{et~al.}(2009)Shen, Narayanaswamy, and Chen]{SPhP2}
Shen,~S.; Narayanaswamy,~A.; Chen,~G. Surface Phonon Polaritons Mediated Energy
  Transfer between Nanoscale Gaps. \emph{Nano Lett.} \textbf{2009}, \emph{9},
  2909--2913\relax
\mciteBstWouldAddEndPuncttrue
\mciteSetBstMidEndSepPunct{\mcitedefaultmidpunct}
{\mcitedefaultendpunct}{\mcitedefaultseppunct}\relax
\EndOfBibitem
\bibitem[Mulet \latin{et~al.}(2002)Mulet, Joulain, Carminati, and
  Greffet]{SPhP3}
Mulet,~J.-P.; Joulain,~K.; Carminati,~R.; Greffet,~J.-J. ENHANCED RADIATIVE
  HEAT TRANSFER AT NANOMETRIC DISTANCES. \emph{Microscale Thermophys. Eng.}
  \textbf{2002}, \emph{6}, 209--222\relax
\mciteBstWouldAddEndPuncttrue
\mciteSetBstMidEndSepPunct{\mcitedefaultmidpunct}
{\mcitedefaultendpunct}{\mcitedefaultseppunct}\relax
\EndOfBibitem
\bibitem[Iizuka and Fan(2015)Iizuka, and Fan]{SPhP4}
Iizuka,~H.; Fan,~S. Analytical treatment of near-field electromagnetic heat
  transfer at the nanoscale. \emph{Phys. Rev. B} \textbf{2015}, \emph{92},
  144307\relax
\mciteBstWouldAddEndPuncttrue
\mciteSetBstMidEndSepPunct{\mcitedefaultmidpunct}
{\mcitedefaultendpunct}{\mcitedefaultseppunct}\relax
\EndOfBibitem
\bibitem[Chiloyan \latin{et~al.}(2015)Chiloyan, Garg, Esfarjani, and
  Chen]{SPhP5}
Chiloyan,~V.; Garg,~J.; Esfarjani,~K.; Chen,~G. Transition from near-field
  thermal radiation to phonon heat conduction at sub-nanometre gaps. \emph{Nat.
  Commun.} \textbf{2015}, \emph{6}, 6755\relax
\mciteBstWouldAddEndPuncttrue
\mciteSetBstMidEndSepPunct{\mcitedefaultmidpunct}
{\mcitedefaultendpunct}{\mcitedefaultseppunct}\relax
\EndOfBibitem
\bibitem[Song \latin{et~al.}(2015)Song, Ganjeh, Sadat, Thompson, Fiorino,
  Fern{\'a}ndez-Hurtado, Feist, Garcia-Vidal, Cuevas, Reddy, and
  Meyhofer]{SPhP6}
Song,~B.; Ganjeh,~Y.; Sadat,~S.; Thompson,~D.; Fiorino,~A.;
  Fern{\'a}ndez-Hurtado,~V.; Feist,~J.; Garcia-Vidal,~F.~J.; Cuevas,~J.~C.;
  Reddy,~P.; Meyhofer,~E. Enhancement of near-field radiative heat transfer
  using polar dielectric thin films. \emph{Nat. Nanotechnol.} \textbf{2015},
  \emph{10}, 253\relax
\mciteBstWouldAddEndPuncttrue
\mciteSetBstMidEndSepPunct{\mcitedefaultmidpunct}
{\mcitedefaultendpunct}{\mcitedefaultseppunct}\relax
\EndOfBibitem
\bibitem[Matsuura \latin{et~al.}(1983)Matsuura, Fukui, and
  Tada]{MATSUURA1983157}
Matsuura,~J.; Fukui,~M.; Tada,~O. {ATR} mode of surface magnon polaritons on
  {YIG}. \emph{Solid State Communications} \textbf{1983}, \emph{45}, 157 --
  160\relax
\mciteBstWouldAddEndPuncttrue
\mciteSetBstMidEndSepPunct{\mcitedefaultmidpunct}
{\mcitedefaultendpunct}{\mcitedefaultseppunct}\relax
\EndOfBibitem
\bibitem[Voss \latin{et~al.}(1985)Voss, Kotzott, and Merten]{disp-FM-85}
Voss,~R.; Kotzott,~R.; Merten,~L. Dispersion of Magnon-Polaritons in Uniaxial
  Ferrites. \emph{physica status solidi (b)} \textbf{1985}, \emph{128},
  159--167\relax
\mciteBstWouldAddEndPuncttrue
\mciteSetBstMidEndSepPunct{\mcitedefaultmidpunct}
{\mcitedefaultendpunct}{\mcitedefaultseppunct}\relax
\EndOfBibitem
\bibitem[Mac\^edo and Camley(2019)Mac\^edo, and Camley]{PhysRevBMac}
Mac\^edo,~R.; Camley,~R.~E. Engineering terahertz surface magnon-polaritons in
  hyperbolic antiferromagnets. \emph{Phys. Rev. B} \textbf{2019}, \emph{99},
  014437\relax
\mciteBstWouldAddEndPuncttrue
\mciteSetBstMidEndSepPunct{\mcitedefaultmidpunct}
{\mcitedefaultendpunct}{\mcitedefaultseppunct}\relax
\EndOfBibitem
\bibitem[Polder(1949)]{Polder49}
Polder,~D. On the theory of ferromagnetic resonance. \emph{The London,
  Edinburgh, and Dublin Philosophical Magazine and Journal of Science}
  \textbf{1949}, \emph{40}, 99--115\relax
\mciteBstWouldAddEndPuncttrue
\mciteSetBstMidEndSepPunct{\mcitedefaultmidpunct}
{\mcitedefaultendpunct}{\mcitedefaultseppunct}\relax
\EndOfBibitem
\bibitem[Miller \latin{et~al.}(2017)Miller, Zhu, and Fan]{Miller4336}
Miller,~D. A.~B.; Zhu,~L.; Fan,~S. Universal modal radiation laws for all
  thermal emitters. \emph{Proceedings of the National Academy of Sciences}
  \textbf{2017}, \emph{114}, 4336--4341\relax
\mciteBstWouldAddEndPuncttrue
\mciteSetBstMidEndSepPunct{\mcitedefaultmidpunct}
{\mcitedefaultendpunct}{\mcitedefaultseppunct}\relax
\EndOfBibitem
\bibitem[Cao \latin{et~al.}(2018)Cao, Fatemi, Fang, Watanabe, Taniguchi,
  Kaxiras, and Jarillo-Herrero]{Unconventional2018}
Cao,~Y.; Fatemi,~V.; Fang,~S.; Watanabe,~K.; Taniguchi,~T.; Kaxiras,~E.;
  Jarillo-Herrero,~P. Unconventional superconductivity in magic-angle graphene
  superlattices. \emph{Nature} \textbf{2018}, \emph{556}, 43--50\relax
\mciteBstWouldAddEndPuncttrue
\mciteSetBstMidEndSepPunct{\mcitedefaultmidpunct}
{\mcitedefaultendpunct}{\mcitedefaultseppunct}\relax
\EndOfBibitem
\bibitem[Cao \latin{et~al.}(2020)Cao, Chowdhury, Rodan-Legrain, Rubies-Bigorda,
  Watanabe, Taniguchi, Senthil, and Jarillo-Herrero]{Strange2020}
Cao,~Y.; Chowdhury,~D.; Rodan-Legrain,~D.; Rubies-Bigorda,~O.; Watanabe,~K.;
  Taniguchi,~T.; Senthil,~T.; Jarillo-Herrero,~P. Strange Metal in Magic-Angle
  Graphene with near Planckian Dissipation. \emph{Phys. Rev. Lett.}
  \textbf{2020}, \emph{124}, 076801\relax
\mciteBstWouldAddEndPuncttrue
\mciteSetBstMidEndSepPunct{\mcitedefaultmidpunct}
{\mcitedefaultendpunct}{\mcitedefaultseppunct}\relax
\EndOfBibitem
\bibitem[Yu \latin{et~al.}(2017)Yu, Liu, Tang, Xu, and Yao]{2017Moir}
Yu,~H.; Liu,~G.~B.; Tang,~J.; Xu,~X.; Yao,~W. Moiré excitons: From
  programmable quantum emitter arrays to spin-orbit–coupled artificial
  lattices. \emph{Science Advances} \textbf{2017}, \emph{3}, e1701696\relax
\mciteBstWouldAddEndPuncttrue
\mciteSetBstMidEndSepPunct{\mcitedefaultmidpunct}
{\mcitedefaultendpunct}{\mcitedefaultseppunct}\relax
\EndOfBibitem
\bibitem[Chen \latin{et~al.}(2019)Chen, Sun, Wang, Gu, Xu, Wu, and
  Gao]{0Direct}
Chen,~W.; Sun,~Z.; Wang,~Z.; Gu,~L.; Xu,~X.; Wu,~S.; Gao,~C. Direct observation
  of van der {W}aals stacking–dependent interlayer magnetism. \emph{Science}
  \textbf{2019}, \emph{366}, 983\relax
\mciteBstWouldAddEndPuncttrue
\mciteSetBstMidEndSepPunct{\mcitedefaultmidpunct}
{\mcitedefaultendpunct}{\mcitedefaultseppunct}\relax
\EndOfBibitem
\bibitem[Sunku \latin{et~al.}(2018)Sunku, Ni, Jiang, Yoo, Sternbach, Mcleod,
  Stauber, Xiong, Taniguchi, and Watanabe]{2018Photonic}
Sunku,~S.~S.; Ni,~G.~X.; Jiang,~B.~Y.; Yoo,~H.; Sternbach,~A.; Mcleod,~A.~S.;
  Stauber,~T.; Xiong,~L.; Taniguchi,~T.; Watanabe,~K.~a. Photonic crystals for
  nano-light in {M}oiré graphene superlattices. \emph{Science} \textbf{2018},
  \emph{362}, 1153--1156\relax
\mciteBstWouldAddEndPuncttrue
\mciteSetBstMidEndSepPunct{\mcitedefaultmidpunct}
{\mcitedefaultendpunct}{\mcitedefaultseppunct}\relax
\EndOfBibitem
\bibitem[Hu \latin{et~al.}(2020)Hu, Krasnok, Mazor, Qiu, and Alù]{2020Moir}
Hu,~G.; Krasnok,~A.; Mazor,~Y.; Qiu,~C.~W.; Alù,~A. Moir\'e Hyperbolic
  Metasurfaces. \emph{Nano Letters} \textbf{2020}, \emph{20}, 3217--3224\relax
\mciteBstWouldAddEndPuncttrue
\mciteSetBstMidEndSepPunct{\mcitedefaultmidpunct}
{\mcitedefaultendpunct}{\mcitedefaultseppunct}\relax
\EndOfBibitem
\bibitem[Gomez-Diaz \latin{et~al.}(2015)Gomez-Diaz, Tymchenko, and
  Alu]{2015Hyperbolic}
Gomez-Diaz,~J.~S.; Tymchenko,~M.; Alu,~A. Hyperbolic Plasmons and Topological
  Transitions Over Uniaxial Metasurfaces. \emph{Physical Review Letters}
  \textbf{2015}, \emph{114}, 233901\relax
\mciteBstWouldAddEndPuncttrue
\mciteSetBstMidEndSepPunct{\mcitedefaultmidpunct}
{\mcitedefaultendpunct}{\mcitedefaultseppunct}\relax
\EndOfBibitem
\bibitem[Hu \latin{et~al.}(2020)Hu, Ou, Si, Wu, and Alù]{Topological2020}
Hu,~G.; Ou,~Q.; Si,~G.; Wu,~Y.; Alù,~A. Topological polaritons and photonic
  magic angles in twisted $\alpha$-{M}o{O}$_3$ bilayers. \emph{Nature}
  \textbf{2020}, \emph{582}, 209--213\relax
\mciteBstWouldAddEndPuncttrue
\mciteSetBstMidEndSepPunct{\mcitedefaultmidpunct}
{\mcitedefaultendpunct}{\mcitedefaultseppunct}\relax
\EndOfBibitem
\bibitem[Wu \latin{et~al.}(2018)Wu, Fu, and Zhang]{2018Influence}
Wu,~X.; Fu,~C.; Zhang,~Z. Influence of h{BN} orientation on the near-field
  radiative heat transfer between graphene/h{BN} heterostructures.
  \emph{Journal of Photonics for Energy} \textbf{2018}, \emph{9}, 1\relax
\mciteBstWouldAddEndPuncttrue
\mciteSetBstMidEndSepPunct{\mcitedefaultmidpunct}
{\mcitedefaultendpunct}{\mcitedefaultseppunct}\relax
\EndOfBibitem
\bibitem[He \latin{et~al.}(2020)He, Qi, Ren, Zhao, and Antezza]{He_20}
He,~M.; Qi,~H.; Ren,~Y.; Zhao,~Y.; Antezza,~M. Magnetoplasmon-surface phonon
  polaritons' coupling effects in radiative heat transfer. \emph{Opt. Lett.}
  \textbf{2020}, \emph{45}, 5148--5151\relax
\mciteBstWouldAddEndPuncttrue
\mciteSetBstMidEndSepPunct{\mcitedefaultmidpunct}
{\mcitedefaultendpunct}{\mcitedefaultseppunct}\relax
\EndOfBibitem
\bibitem[Wu \latin{et~al.}(2020)Wu, Fu, and Zhang]{2020Near}
Wu,~X.; Fu,~C.; Zhang,~Z. Near-Field Radiative Heat Transfer Between Two
  $\alpha$-{M}o{O}$_3$ Biaxial Crystals. \emph{Journal of Heat Transfer}
  \textbf{2020}, \emph{142}, 1--10\relax
\mciteBstWouldAddEndPuncttrue
\mciteSetBstMidEndSepPunct{\mcitedefaultmidpunct}
{\mcitedefaultendpunct}{\mcitedefaultseppunct}\relax
\EndOfBibitem
\bibitem[Luo \latin{et~al.}(2020)Luo, Zhao, and Antezza]{2020Near_field}
Luo,~M.; Zhao,~J.; Antezza,~M. Near-field radiative heat transfer between
  twisted nanoparticle gratings. \emph{Applied Physics Letters} \textbf{2020},
  \emph{117}, 053901\relax
\mciteBstWouldAddEndPuncttrue
\mciteSetBstMidEndSepPunct{\mcitedefaultmidpunct}
{\mcitedefaultendpunct}{\mcitedefaultseppunct}\relax
\EndOfBibitem
\bibitem[Zhou \latin{et~al.}(2020)Zhou, Wu, Zhang, Yi, and
  Mauro]{Polariton_2020}
Zhou,~C.-L.; Wu,~X.-H.; Zhang,~Y.; Yi,~H.-L.; Mauro,~A. Polariton Topological
  Transition Effects on Radiative Heat Transfer. \emph{arXiv:2011.02263}
  \textbf{2020}, \relax
\mciteBstWouldAddEndPunctfalse
\mciteSetBstMidEndSepPunct{\mcitedefaultmidpunct}
{}{\mcitedefaultseppunct}\relax
\EndOfBibitem
\bibitem[Pochi and Yeh(1980)Pochi, and Yeh]{Pochi1980Optics}
Pochi,; Yeh, Optics of anisotropic layered media: A new 4 x 4 matrix algebra.
  \emph{Surface Science Letters} \textbf{1980}, \emph{96}, 4153\relax
\mciteBstWouldAddEndPuncttrue
\mciteSetBstMidEndSepPunct{\mcitedefaultmidpunct}
{\mcitedefaultendpunct}{\mcitedefaultseppunct}\relax
\EndOfBibitem
\bibitem[SM()]{SM}
{See Supplemental Material for derivations and additional details.}\relax
\mciteBstWouldAddEndPunctfalse
\mciteSetBstMidEndSepPunct{\mcitedefaultmidpunct}
{}{\mcitedefaultseppunct}\relax
\EndOfBibitem
\bibitem[Yu \latin{et~al.}(2017)Yu, Sun, and Gao]{YIG16}
Yu,~W.; Sun,~H.; Gao,~L. Magnetic control of {G}oos-{H}{\"a}nchen shifts in a
  yttrium-iron-garnet film. \emph{Sci. Rep.} \textbf{2017}, \emph{7},
  45866\relax
\mciteBstWouldAddEndPuncttrue
\mciteSetBstMidEndSepPunct{\mcitedefaultmidpunct}
{\mcitedefaultendpunct}{\mcitedefaultseppunct}\relax
\EndOfBibitem
\bibitem[Mac\^edo \latin{et~al.}(2020)Mac\^edo, C.~Holland, G.~Baity,
  L.~Livesey, L.~Stamps, P.~Weides, and A.~Bozhko]{Rair-Macdo2020}
Mac\^edo,~R.; C.~Holland,~R.; G.~Baity,~P.; L.~Livesey,~K.; L.~Stamps,~R.;
  P.~Weides,~M.; A.~Bozhko,~D. An Electromagnetic Approach to Cavity
  Spintronics. \emph{arXiv:2007.11483} \textbf{2020}, \relax
\mciteBstWouldAddEndPunctfalse
\mciteSetBstMidEndSepPunct{\mcitedefaultmidpunct}
{}{\mcitedefaultseppunct}\relax
\EndOfBibitem
\bibitem[Boventer \latin{et~al.}(2018)Boventer, Pfirrmann, Krause, Sch\"on,
  Kl\"aui, and Weides]{Boventer2018}
Boventer,~I.; Pfirrmann,~M.; Krause,~J.; Sch\"on,~Y.; Kl\"aui,~M.; Weides,~M.
  Complex temperature dependence of coupling and dissipation of cavity magnon
  polaritons from millikelvin to room temperature. \emph{Phys. Rev. B}
  \textbf{2018}, \emph{97}, 184420\relax
\mciteBstWouldAddEndPuncttrue
\mciteSetBstMidEndSepPunct{\mcitedefaultmidpunct}
{\mcitedefaultendpunct}{\mcitedefaultseppunct}\relax
\EndOfBibitem
\bibitem[Zhang(2020)]{ZhangMing}
Zhang,~Z.~M. \emph{Nano/Microscale Heat Transfer 2nd edition}; Springer, Cham,
  2020\relax
\mciteBstWouldAddEndPuncttrue
\mciteSetBstMidEndSepPunct{\mcitedefaultmidpunct}
{\mcitedefaultendpunct}{\mcitedefaultseppunct}\relax
\EndOfBibitem
\bibitem[Latella and Ben-Abdallah(2017)Latella, and Ben-Abdallah]{GTM2017}
Latella,~I.; Ben-Abdallah,~P. Giant Thermal Magnetoresistance in Plasmonic
  Structures. \emph{Phys. Rev. Lett.} \textbf{2017}, \emph{118}, 173902\relax
\mciteBstWouldAddEndPuncttrue
\mciteSetBstMidEndSepPunct{\mcitedefaultmidpunct}
{\mcitedefaultendpunct}{\mcitedefaultseppunct}\relax
\EndOfBibitem
\bibitem[Iizuka and Fan(2018)Iizuka, and Fan]{Iizuka2018}
Iizuka,~H.; Fan,~S. Significant Enhancement of Near-Field Electromagnetic Heat
  Transfer in a Multilayer Structure through Multiple Surface-States Coupling.
  \emph{Phys. Rev. Lett.} \textbf{2018}, \emph{120}, 063901\relax
\mciteBstWouldAddEndPuncttrue
\mciteSetBstMidEndSepPunct{\mcitedefaultmidpunct}
{\mcitedefaultendpunct}{\mcitedefaultseppunct}\relax
\EndOfBibitem
\end{mcitethebibliography}

\end{document}